\documentstyle[aaspp4,psfig]{article}

\newcommand{\Alfven}{ Alfv\'{e}n }

\newcommand{\nn}{\mbox{} \nonumber \\ \mbox{} &&}
\newcommand{\ob}{\omega_B}
\newcommand{\op}{\omega_p}
\newcommand{\om}{\omega}

\newcommand{\be}{\begin{equation}}
\newcommand{\ee}{\end{equation}}
\newcommand{\ba}{\begin{eqnarray}}
\newcommand{\ea}{\end{eqnarray}}

\begin{document}
\title {Depolarization of  Pulsar Radio Emission}
\author{Maxim Lyutikov}
\affil{Canadian Institute for Theoretical Astrophysics, 60 St. George, Toronto, Ont,  M5S 3H8, Canada}
\date {draft}

\begin{abstract}
We
 show that intensity dependent depolarization
of single pulses (e.g. Xiluoris
et al. 1994) may be due to the
nonlinear decay of the "upper" ordinary mode
into an unpolarized extraordinary
mode and a backward propagating wave.
The decay occurs in the innermost parts
 of the pulsar magnetosphere for obliquely
propagating O waves.
\end{abstract}


\section{Introduction}
There is a marked anticorrelation between
 the intensity of a pulse and a degree
of linear polarization  (Manchester et al.
1975, Xilouris et al. 1994). Depolarization
may be due to either a generic feature of
the  emission mechanism or wave
propagation in the pulsar magnetosphere.
In the latter case two mechanisms
 are possible - linear and nonlinear coupling.
Linear coupling is due to the breakdown of the
geometric  optics approximation for polarization
changes (Zheleznyakov 1996).
The efficiency of linear coupling is
 independent of the wave amplitude.
Nonlinear coupling is due to the wave-wave
interaction processes. Its
 efficiency is proportional to the amplitudes
or intensities of the interacting waves.
The observed inverse  dependence of
the degree of linear polarization on intensity
suggests that it is the nonlinear mechanism that
 is responsible for depolarization.

 The pair plasma of pulsar magnetospheres supports
 three types of linearly polarized waves:
 transverse  extraordinary (X)
mode, and mixed \Alfven
(A) and ordinary (O) modes. For
 a given wave vector
 the frequencies of the waves satisfy
 $\om_O > \om_X > \om_A$, so that the possible induced decays
are $  O \rightarrow 2 X $,  $  O \rightarrow X+A$, $X\rightarrow A+ X$.
In all cases the product
waves propagate in approximately opposite
 directions. 
\footnote{Decay of X into two forward propagating X modes is suppressed by magnetic field -
the matrix element $\sim 1/\om_B^2$, compare with (Eq. \ref{V}).}

\begin{figure}
\psfig{file=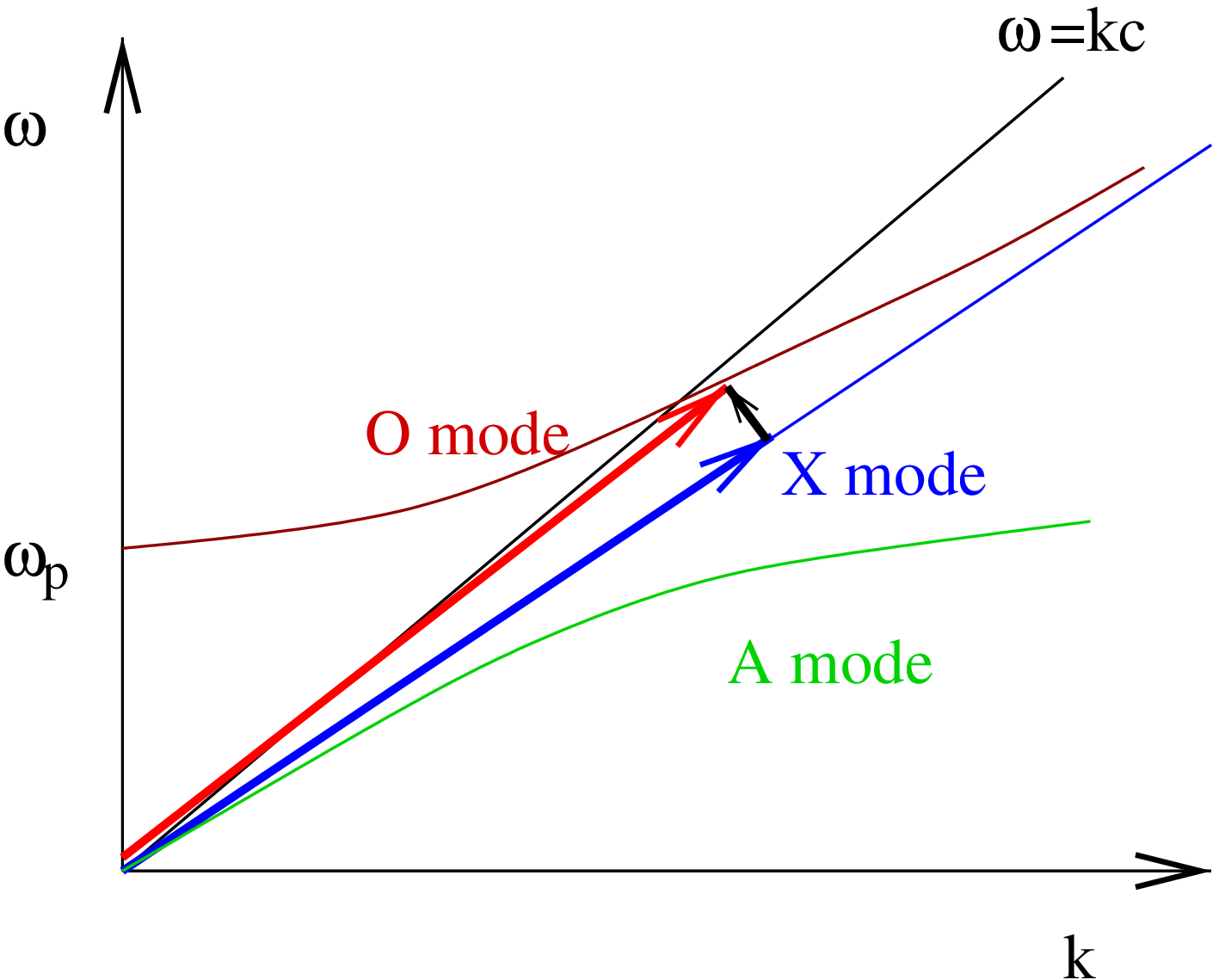,width=12.0cm}
\caption{
Decay of O mode into X mode plus backward propagating  X or \Alfven mode. }
\label{fig}
\end{figure}

Suppose
that the emission mechanism generates waves
of the ordinary (O) mode (some emission
 mechanisms do favor the O mode for
the generation of radio emission). Initially
the O mode may be  strongly polarized.
The preferred polarization of the emitted O
mode may be related to the osculating
 plane of the curved magnetic field lines:
the emission mechanism may generate waves
polarized mostly in  the osculating plane
(e.g. curvature emission) or perpendicular
 to the this plane (Cherenkov-drift emission  (Lyutikov et al.
1999)). As the emitted O mode propagates in
the magnetosphere, it will experience 
induced decay into X modes and backward
propagating waves. In the leading order of
expansion in $1/\om_B$
and for small angles of propagation with respect
to  the magnetic field, the decay of the O mode
will produce  two orthogonally polarized 
modes with {\it  random polarization}
 with respect to the initial O mode (appendix A).
  The rate of this decay is  proportional to  the intensity  of the wave
 (the rate of decay
is proportional to the wave amplitude
 for the spectral narrow line, and to the wave  intensity  for the broad
spectrum).
The product X modes can in turn decay into \Alfven waves. But
\Alfven modes cannot escape from plasma and are
damped  by the  Cherenkov interaction with plasma particles (Arons \& Barnard 1986).
Thus,  the induced decay of $X \rightarrow A + X $
(with the X mode propagating backwards) is of no interest to us  since it  generates
unobservable waves \footnote{\Alfven waves may be converted back to the X or O mode
due to linear  coupling in a strongly turbulent plasma. In any case the resulting waves
will be unpolarized.}.

The processes that will contribute to depolarization
  include decays (Fig. \ref{fig}) $  O \rightarrow 2 X $ and
$  O \rightarrow X+A$ (in the latter  case it is the \Alfven wave that is propagating backwards),
and  inverse 
 processes of coalescence of backward propagating X or \Alfven mode with a
forward propagating X mode.
The  inverse 
 processes will be proportional to the product of the intensities of   coalescing  waves.
In some case this may be an important process, but the convection of backward propagating waves
from  the interaction region will most probably reduce their intensity considerably making the
coalescence ineffective.
If inverse  processes are ineffective, virtually all of the O mode may be converted
into the X mode; the X modes may then be 
 as intense as the initial O mode (backward propagating waves
carry off only a small fraction of  energy).

\section{Induced decay of O mode}

For the three wave processes to take place, the participating waves
 should satisfy resonance conditions (conservation of energy and
momentum along the field)
\begin{eqnarray}
&&
\om_o=\om_X+\om_2
\nn
{\bf k}_{O}= {\bf k} _{X}+  {\bf k}_2
\label{res}
\end{eqnarray}
and more subtle
conditions on the polarization determined by the
matrix elements $A ( {\bf k}_0,  {\bf k}_X,  {\bf k}_2) $
of the  third order nonlinear current (Melrose
1978 Eqs. (10.105) and (10.125)). 
The   matrix element for the decay of the O mode into the  X mode and another transverse mode 
(which could be X or \Alfven wave) 
 are calculated in  appendix A. 
A typical growth rate for the decay of the O mode is (Eq. \ref{q})
\be 
\Gamma \approx  { q^2 \over  m ^2 c^3} { \op^6   \sin ^4 \theta \over  \gamma_p^3 \ob^2 \om^4 }
{W_k }
\ee
where $W_k$ is the wave intensity per unit interval of wave vectors $k$,
$q $ is the elementary charge, $m$ is a mass of an electron,
$c$ is a velocity of light,
$\om_p$  is a plasma frequency, $\theta$ is an angle of propagation with
 respect to  the magnetic field,  $\om_B$ is a cyclotron frequency, and
$\gamma_p$ is a Lorentz factor of the plasma motion.

Wave intensity $W_k$  can be related to the observed flux $F_{\nu} $: 
$ W_k= {1\over 2 \pi} F_{\nu} \left( {d \over r} \right)^2 $ 
where $d/r$ is the ratio of the
distance to the pulsar to the typical radius where transformations occur. If we parametrize the
plasma density in terms of the Goldreich-Julian density 
$\op^2 =2 \lambda_M \Omega \ob$ (where $\lambda_M \approx 10^6$ is a multiplicity factor
that gives a number of secondary pairs per primary particle)  then we can express the growth
rate in terms of  the observed flux and pulsar  rotation frequency $\Omega$
\be 
\Gamma \approx  { q^2 \over  m ^2 c^3} {  \lambda_M ^3 \Omega^3 \om_{B, NS}   \sin ^4 \theta
 \over  \gamma_p^3  \om^4 } \, \left( { R_{NS} \over r} \right)^3 \,
  F_{\nu}  \left( {d \over r} \right)^2
\label{q1q}
\ee
where $\om_{B, NS} $ is the surface cyclotron frequency and $R_{NS}$ is the radius of the neutron star.

For a pulsar at $d= 1 \, {\rm kpc}$, with rotation frequency $\Omega=50 \, {\rm rad/sec}$,
average Lorentz factor of plasma $\gamma_p=10$, observation frequency $\nu = 1$ GHz,
typical interaction radius $r= 5 \times 10^6 \, {\rm cm}$,  
and a peak flux of 1 Jy, we find the growth
rate of the decay instability
\be
 \Gamma \approx 10^3  \sin ^4 \theta {\rm sec}^{-1}
\label{qq}
\ee
This growth rate may be called marginal:
efficient wave transformation occurs only for
comparatively large angles of propagation $\theta \geq 0.1$ and it is  quite sensitive to the
choice of plasma parameters: changing  the plasma density or the streaming Lorentz factor
may make  the growth time  larger that the pulsar's period - making the  wave transformation
 ineffective.

For a given intensity, the  growth rate (\ref{q1q}) decreases with radius $\propto r^{-3}$
(for a given observed flux growth rate $\propto r^{-5}$ ); thus,
the most effective transformation occurs deep inside the pulsar magnetosphere.
The effectiveness of the wave  transformation in different pulsars depends mostly
on the plasma production ($\lambda_M $ and $\gamma_p$). Denser plasma streaming with a
smaller Lorentz factor is more efficient for wave conversion.
If the plasma production is independent of the
pulsar period, then the wave  transformation should
be more efficient in the short period pulsars.

The frequency dependence  of the wave transformation may also be complicated. On one
hand, the efficiency of transformation (Eq. \ref{q1q}) is inversely proportional to the frequency,
and the pulsar's spectral $F_{\nu}$ are usually power laws with index $-1.6$
(Lorimer et al. 1995).
 On the other hand, higher frequencies are  believed to be emitted lower in the magnetosphere, so
they acquire a larger angle with respect to  the magnetic field in the interaction region;
 thus,  they may have a larger decay rate. The lower frequencies,
which are emitted at  heights larger than $\sim$ ten stellar radii do not
acquire large angles before they leave the region of effective wave transformation.

\section{Conclusion}

We showed that induced decay of the "upper" O mode into the  unpolarized X mode
 in the inner parts of the  pulsar magnetosphere
may explain observed intensity dependent depolarization.
When the intensity of the initial O mode is not high enough  or the angle of propagation
with respect to the magnetic field is smalls,  the decay is weak and  the 
original O mode is observed. When the intensity of the  O mode is high and the wave is 
propagating at a considerable angle  with respect to the  magnetic field,   it
may completely decay into
 X modes (and backward propagating waves).
Generally, the growth rate of the decay instability
is marginal - variations in the plasma parameters and the  structure of the magnetic field
may make it  either ineffective or extremely effective.

\acknowledgements
I would like to thank Anuj Parikh for his comments on the manuscript.

\appendix

\section{Calculation of matrix element}

Consider a decay of a high frequency ($\om _O \gg \om_P$)
O mode into an X mode and another mode, unspecified at this point.
Let the wave vectors and polarizations be
\begin{eqnarray}
&
{\bf k}_0 = k_0
 \{ \sin \theta,0,\cos \theta\} , &
{\bf e} _O = \{  \cos \theta,0,  \sin \theta\} 
\nonumber \\
&
 {\bf k}_X =k_X \{ \cos ({\phi_X})\,\sin (\theta _X),
   \sin ({\phi_X})\,\sin (\theta _X),\cos (\theta _X)\} ,
&
{\bf e}_X = \{   \sin ({\phi_X}) ,   \sin ({\phi_X}), 0 \}
\nonumber \\
&
  {\bf k}_2 = k_2 
 \{ \cos \phi_2\,\sin (\theta _2),
   \sin \phi_2\,\sin (\theta _2),\cos (\theta _2)\} , & 
{\bf e}_2 =
\{ e_{2x} ,e_{2y} ,e_{2z} \}
\end{eqnarray}
where $\theta$  and $\phi$  are polar coordinates in a frame aligned with the 
magnetic field, which is assumed to be in the  $z$ direction.

In a cold magnetized  plasma the nonlinear third order currents (which determine the 
3-wave interactions) are 
given by  Melrose 1978 Eqs. (10.105) and (10.125). 
Two simplifications are possible in the case of strong magnetized electron-positron plasma.
First, nonlinear current terms which are proportional to an odd power of the sign of the charge
will cancel out. Secondly, we can make an expansion in powers  $1/\ob$ 
and keep the lowest order. Under these assumption 
the matrix element is 
\be
V=
\bar{S}_{ijl} e_{0,i}^{\ast} e_{X,j}  e_{2,l}  \approx -i
 { e \over m c } { \op^2 \sin \theta \over  \om \, \ob } 
\sin (\phi_X - \phi_2)
\label{V}
\ee
where we assumed that the product waves propagate approximately along the field line,
and wave 2 propagates backward.

Two conclusions may be made on analyzing 
Eq. (\ref{V}). First, for the decay to occur, the initial O mode
should be obliquely propagating ($\theta \neq 0$). 
Secondly,
the polarizations of the product wave are completely random with respect to the
initial O mode and approximately  orthogonal to each other.
Thus, the decay of the linearly polarized O mode produces unpolarized waves.
Microscopically, this is due to the fact that it is the $z$ component of the
initial  O mode that couples to the product modes. A more physical description of the
3 wave interaction in pair plasma  may be found in Machabeli \& Rogava 1983.

Given the matrix element (\ref{V}) we can find the the probability of emission
in the random phase approximation (which assumes that radiation is broadband)
 (Melrose 1978)
\be
u = 4 (2 \pi)^7 V^2 {  \omega_X \omega_2  \over \omega } \delta({\bf k} _0 -
{\bf k} _X - {\bf k} _2) \delta(\omega_0 - \om_X -\om_2).
\ee
Then 
the characteristic nonlinear decay  time  is
\be 
\Gamma \approx  { q^2 \over  m ^2 c^3} { \op^6   \sin ^4 \theta \over  \gamma_p^3 \ob^2 \om^4 }
{W_k },
\label{q}
\ee
where we took into
account that $\omega_2 = \omega_O - \omega_X \approx { \op^2 \sin ^2 \theta \over \gamma_p^3  \om } $.
\end{document}